\documentclass{webofc}

\usepackage[varg]{txfonts}   
\usepackage{hyperref}
\usepackage{url}

\hypersetup{colorlinks=true,citecolor=blue,urlcolor=blue,linkcolor=blue}
\begin{document}
\title{The impact of glasma on heavy quark spectra \\ and correlations}
%
%

\author{
\firstname{Dana} \lastname{Avramescu}\inst{1,2}\fnsep\thanks{\email{dana.d.avramescu@jyu.fi}} 
\and
\firstname{Vincenzo} \lastname{Greco}\inst{3,4}\fnsep\thanks{\email{greco@lns.infn.it}} 
\and
\firstname{Tuomas} \lastname{Lappi}\inst{1,2}\fnsep\thanks{\email{tuomas.v.v.lappi@jyu.fi}}
\and
\firstname{Heikki} \lastname{M\"{a}ntysaari}\inst{1,2}\fnsep\thanks{\email{heikki.mantysaari@jyu.fi}}
\and
\firstname{David} \lastname{M\"{u}ller}\inst{5}\fnsep\thanks{\email{dmueller@hep.itp.tuwien.ac.at}}
}

\institute{
Department of Physics, University of Jyväskylä,  P.O. Box 35, 40014 University of Jyväskylä, Finland
\and
Helsinki Institute of Physics, P.O. Box 64, 00014 University of Helsinki, Finland 
\and
Department of Physics and Astronomy, University of Catania, Via S. Sofia 64, I-95123 Catania, Italy
\and
INFN-Laboratori Nazionali del Sud, Via S. Sofia 62, I-95123 Catania, Italy
\and 
Institute for Theoretical Physics, TU Wien, Wiedner Hauptstraße 8, A-1040 Vienna, Austria
}

\abstract{We investigate the effect of the glasma classical color fields, produced in the very early stage of heavy-ion collisions, on the transport of heavy quarks. The glasma fields evolve according to the classical Yang-Mills equations, while the dynamics of heavy quarks is described by Wong's equations. We numerically solve these equations and compute the transport coefficient $\kappa$, which is anisotropic and initially very large. Further, we extract observables sensitive to the initial glasma stage. The heavy quark nuclear modification factor $R_{AA}$ is affected by the glasma but the effect is moderate compared to the nPDF contribution. Our main finding is that the glasma has a large impact on the azimuthal correlation between $Q\overline{Q}$ pairs, initially produced back-to-back.}
\maketitle

\section{Introduction}

The theoretical description of heavy-ion collisions requires dividing the collision process into distinct stages. In a weak coupling picture \cite{Berges:2020fwq}, the earliest stage is the glasma \cite{Lappi:2006hq}. It comes from high-energy collisions within the Color Glass Condensate (CGC) framework \cite{Gelis:2010nm}. The collision produces strong classical gluon fields and enables the study of out-of-equilibrium matter using non-perturbative methods. Recently, there has been increasing interest in studying the effect of the pre-equilibrium stages on the transport of heavy quarks and jets \cite{Andres:2019eus,Sun:2019fud,Ipp:2020nfu,Carrington:2021dvw,Avramescu:2023qvv,Boguslavski:2023fdm,Boguslavski:2023alu,Barata:2024xwy,Oliva:2024rex}. In this work, we focus on how the glasma affects heavy quark observables.

\section{Heavy quark transport in glasma}

\noindent\textit{\textbf{Glasma.}} The glasma fields are derived using CGC, an effective field theory for high-energy QCD. At high energies, a nucleus mostly contains soft gluons that carry a small fraction of the total momentum. The CGC separates the soft and hard partons: the hard partons produce a color current $J^\mu$, which generates the gauge field $A^\mu$ of the soft gluons. The gluon fields have large occupation numbers and are treated as classical. These fields obey the corresponding classical Yang-Mills field equations
\begin{equation}
    \mathcal{D}_\mu F^{\mu\nu}=J^\nu,
\end{equation}
with $\mathcal{D}_\mu=\partial_\mu-\mathrm{i}g A_\mu$ and $F^{\mu\nu}=\partial^\mu A^\nu-\partial^\nu A^\mu-\mathrm{i}g[A^\mu, A^\nu]$. The color current $J^\mu$ is generated by the color charge density within the nucleus. At high energy, this color charge is confined to a thin sheet in the transverse plane $\boldsymbol{x}_T$ and we use the McLerran-Venugopalan (MV) model to describe it, namely $\langle \rho^a(\boldsymbol{x}_T)\,\rho^b(\boldsymbol{y}_T)\rangle = (g^2\mu)^2\delta^{ab}\delta^{(2)}(\boldsymbol{x}_T-\boldsymbol{y}_T)$, with $g^2\mu\propto Q_s$. The saturation momentum $Q_s$, which is proportional to the color charge density, is the only physical parameter. In our work, we set $Q_s=2\,\mathrm{GeV}$, appropriate for central LHC collisions. 

Using the CGC theory and the MV model, one obtains analytical expressions for the gluon fields produced by a single high-energy nucleus before the collision. The glasma is obtained in the collisions of such nuclei and the glasma field can be numerically solved assuming boost invariance, namely that the fields are $\eta$-independent. Initially, the resulting glasma consists of longitudinal electric and magnetic fields extending between the colliding nuclei. These fields form color flux tubes, a key feature of the glasma. The electric and magnetic energy densities are initially large but then decay, and are correlated within color flux tubes \cite{Lappi:2006hq}. These unique features can be studied using heavy quarks and jets as probes.

\begin{figure}
\centering
\sidecaption
\includegraphics[width=6cm,clip]{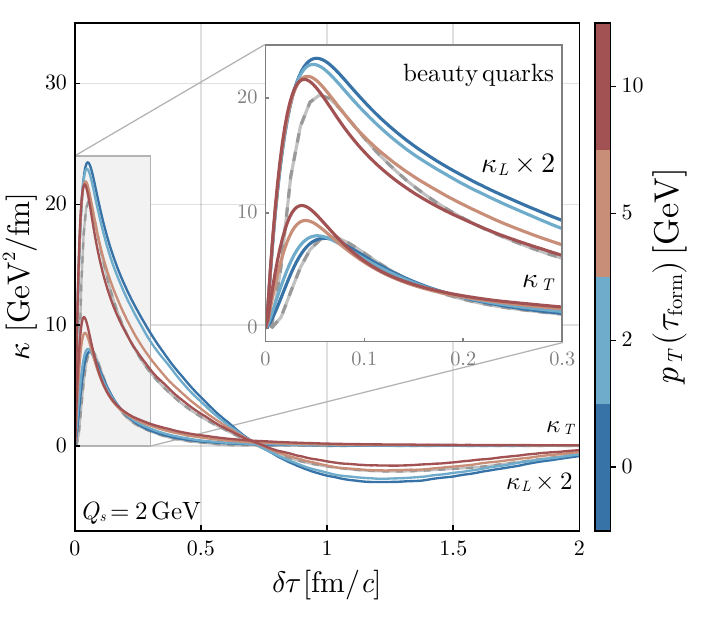}
\caption{The transport coefficient $\kappa$ as a function of the relative proper time $\delta\tau\equiv \tau-\tau_\mathrm{form}$ for beauty quarks, with $\tau_\mathrm{form}=1/2m$. The \textit{gray line} corresponds to infinitely massive quarks, while the \textit{colored lines} correspond to beauty quarks initialized with different $p_T(\tau_\mathrm{form})$ values. The longitudinal $\kappa_L$ and transverse $\kappa_T$ contributions are shown separately, with $\kappa_L>\kappa_T$ in the early times. The heavy quarks evolve in a glasma with saturation momentum $Q_s=2\,\mathrm{GeV}$.}
\label{fig:kappa}   
\end{figure}

\noindent\textit{\textbf{Heavy quarks in glasma.}} In our study \cite{Avramescu:2023qvv} we developed a framework to simulate the trajectories of beauty and charm quarks in the glasma fields. For this, we use classical equations of motion, specifically Wong’s equations. These describe the evolution of the position $x^\mu$, momentum $p^\mu$, and color charge $Q$ for a classical particle with mass $m$ in a classical background gauge field $A^\mu$ as
\begin{equation}
    \frac{\mathrm{d}x^\mu}{\mathrm{d}\tau}=\frac{p^\mu}{m},\quad \dfrac{\mathrm{d}p^\mu}{\mathrm{d}\tau}=\frac{g}{T_R}\mathrm{Tr}\{QF^{\mu\nu}\}\frac{p_\nu}{m},\quad \frac{\mathrm{d}Q}{\mathrm{d}\tau}=-\mathrm{i}g[A_\mu,Q]\frac{p^\mu}{m},
\end{equation}
where $\tau$ is the relativistic proper-time and $\mathrm{Tr}\{T^aT^b\}=T_R\delta^{ab}$. The quarks experience momentum deflections due to the color Lorentz force from the glasma’s electric and magnetic fields, while the quark’s color charge rotates in the SU($3$) color space.

We use our solver to calculate the heavy quark transport coefficient $\kappa$ in the glasma, see Fig.~\ref{fig:kappa}. This involves extracting the momentum broadening $\delta p^2_i(\tau)=p_i^2(\tau)-p_i^2(\tau_\mathrm{form})$, averaging over glasma events and particle trajectories, then taking the Milne proper time derivative $\kappa_i(\tau)=\mathrm{d}\langle \delta p_i^2(\tau)\rangle/\mathrm{d}\tau$, where $i=L,T$ refers to the longitudinal and transverse directions with respect to the beam axis. We find that $\kappa$ in the glasma is anisotropic, with larger values along the longitudinal than transverse direction at early times. The $\kappa$ in glasma exhibits a large initial peak, indicating a strong glasma impact on heavy quark transport. 

\section{Heavy quark observables}

Since $\kappa$ is not a directly measurable quantity, we investigate observables that are potentially sensitive to the glasma stage. In our work, we consider the nuclear modification factor and heavy quark pair correlations in the glasma \cite{Avramescu:2024xts,Avramescu:2024poa}. We only examine the glasma stage without coupling it to later stages.

\begin{figure*}
\centering
\vspace*{1cm}      
\includegraphics[width=0.8\textwidth,clip]{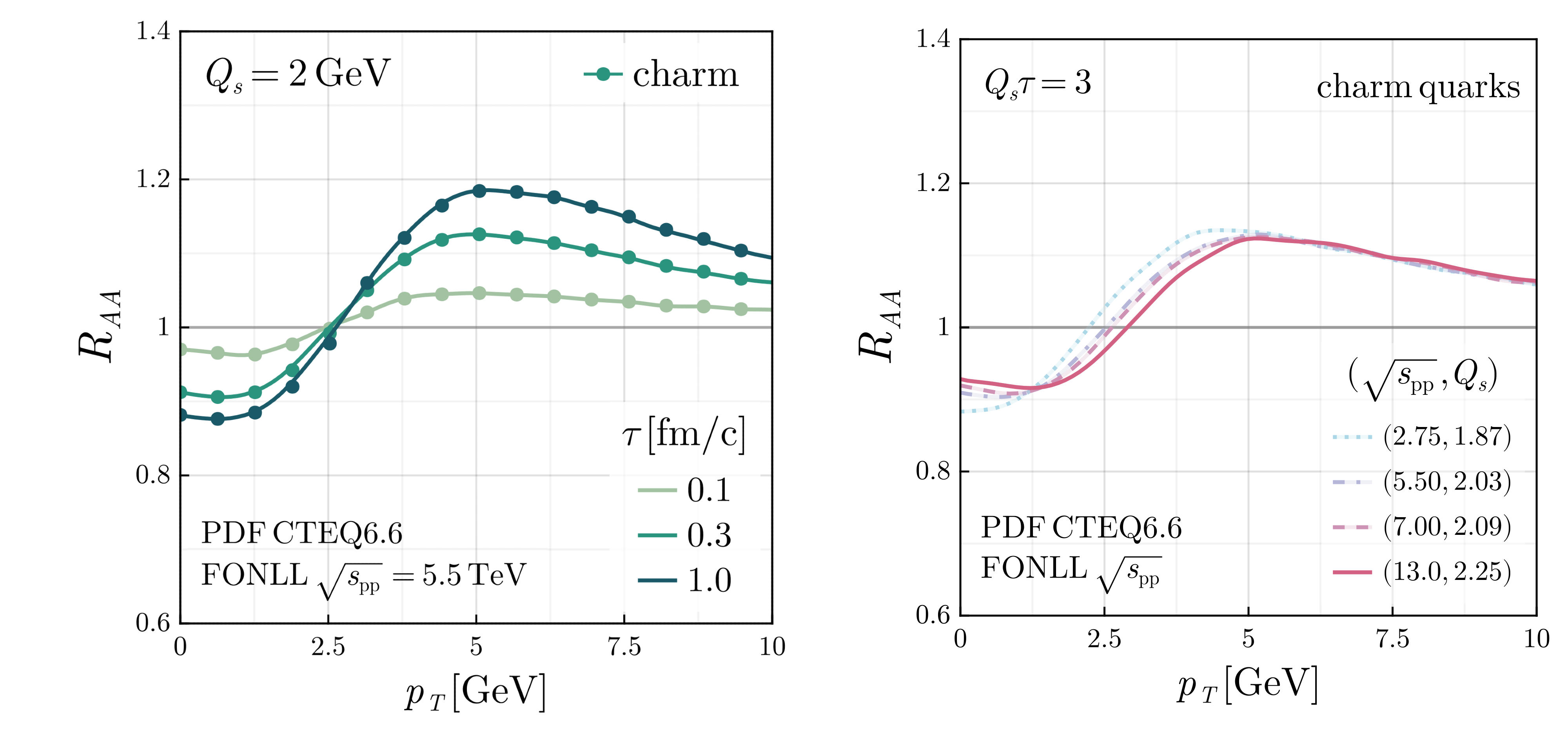}
\caption{The proper time dependence of $R_{AA}(p_T)$ at various $\tau$ values for fixed center-of-mass energy $\sqrt{s}$ and saturation momentum $Q_s$ (\textit{left panel}) and the combined $(\sqrt{s},Q_s)$ dependence for fixed $Q_s\tau$ (\textit{right panel}) for charm quarks.}
\label{fig:raa_tau_energy_dep}       
\end{figure*}

\begin{figure}
\centering
\sidecaption
\includegraphics[width=5cm,clip]{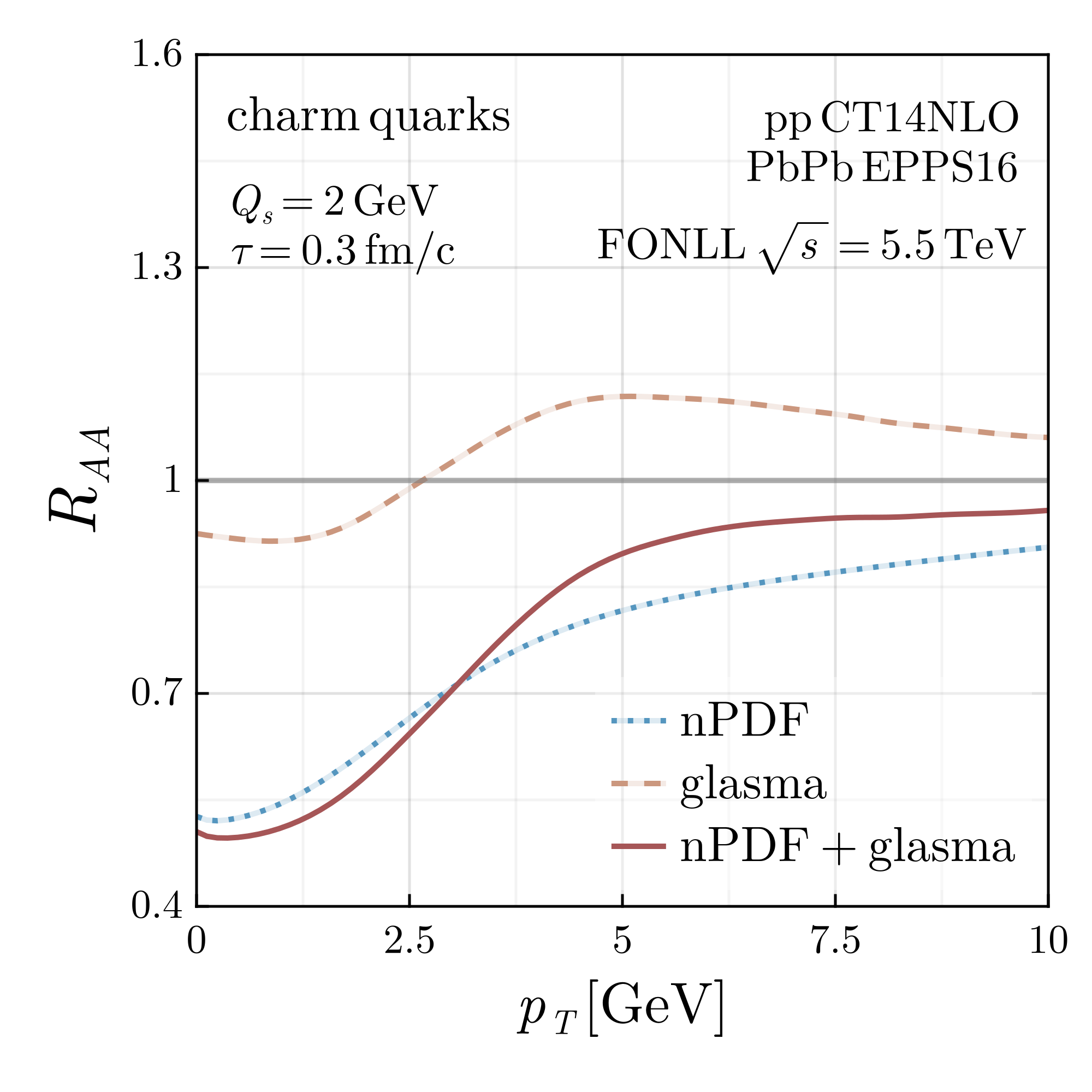}
\caption{$R_{AA}(p_T)$ in glasma without nPDFs (\textit{dashed red curve}), compared with a nPDF-modified $R_{AA}$ using EPPS16 (\textit{red curve}) and the contribution with only the nPDF effect (\textit{blue curve}). The results are shown for charm quarks at $\tau=0.3\,\mathrm{fm/c}$, the typical switching time to kinetic theory or hydrodynamics, at $Q_s=2\,\mathrm{GeV}$, roughly corresponding to the collision energy $\sqrt{s}=5.5\,\mathrm{TeV}$ used in the FONLL initial heavy-quark spectra.}
\label{fig:raa}   
\end{figure}

\noindent\textit{\textbf{Nuclear modification factor.}} We first investigate the nuclear modification factor, obtained from $p_T$ spectra. Since the glasma causes $p_T$ broadening, a flat $p_T$ distribution gets shifted by the glasma fields from small to moderate $p_T$ values and then reaches a plateau at higher $p_T$. To study realistic heavy quark spectra, we use input from the Fixed-Order+Next-to-Leading
Logarithm (FONLL) \cite{Cacciari:1998it} heavy quark production calculation, which gives the $p_T$ spectrum for bare heavy quarks. We then evolve this spectrum through the glasma stage.

To measure the glasma’s effect on spectra, we calculate the nuclear modification factor $R_{AA}$. This is roughly the ratio of the $AA$ to $pp$ spectra, more precisely
\begin{equation}
    R_{AA}(\tau)=\frac{1}{A^2}\frac{\sigma_\mathrm{tot}^{AA}}{\sigma_\mathrm{tot}^{pp}}\frac{\mathrm{d}N^{AA}/\mathrm{d}p_T(\tau)}{\mathrm{d}N^{pp}/\mathrm{d}p_T}.
\end{equation}
In our setup, the $AA$ spectrum evolves in the glasma, while the $pp$ cross-section comes from the FONLL calculation. We assume no glasma in $pp$ collisions. At low $p_T$, $R_{AA}<1$ shows suppression due to $p_T$ migration to higher values. This suppression is balanced by an enhancement at higher $p_T$, keeping the total number of quarks unchanged. Further, we study how $R_{AA}$ changes over time and conclude that the shape of $R_{AA}$ gets more pronounced with temporal evolution. Additionally, we explore the energy dependence of $R_{AA}$. The two energy scales we use are the saturation momentum $Q_s$ and the collision energy entering the FONLL calculation $\sqrt{s}$. For charm quarks, simultaneously increasing $\sqrt{s}$ and $Q_s$ changes $R_{AA}$ at low to intermediate $p_T$, although the effect is mild. All these results are collected in Fig.~\ref{fig:raa_tau_energy_dep}.

Lastly, we compare the glasma's effect on $R_{AA}$ with the initial state contribution coming from the nuclear parton distribution functions (nPDFs). The nPDFs describe how parton densities in nuclei differ from those in protons. Specifically, gluon shadowing in nPDFs reduces the gluon density at small-$x$, further suppressing heavy quark production. In our work, we use EPPS16 \cite{Eskola:2016oht} in the FONLL calculation. The main finding shown in Fig.~\ref{fig:raa} is that while the glasma has an effect, nPDFs dominate the suppression of $R_{AA}$. 

\begin{figure*}
\centering
\vspace*{1cm}      
\includegraphics[width=\textwidth,clip]{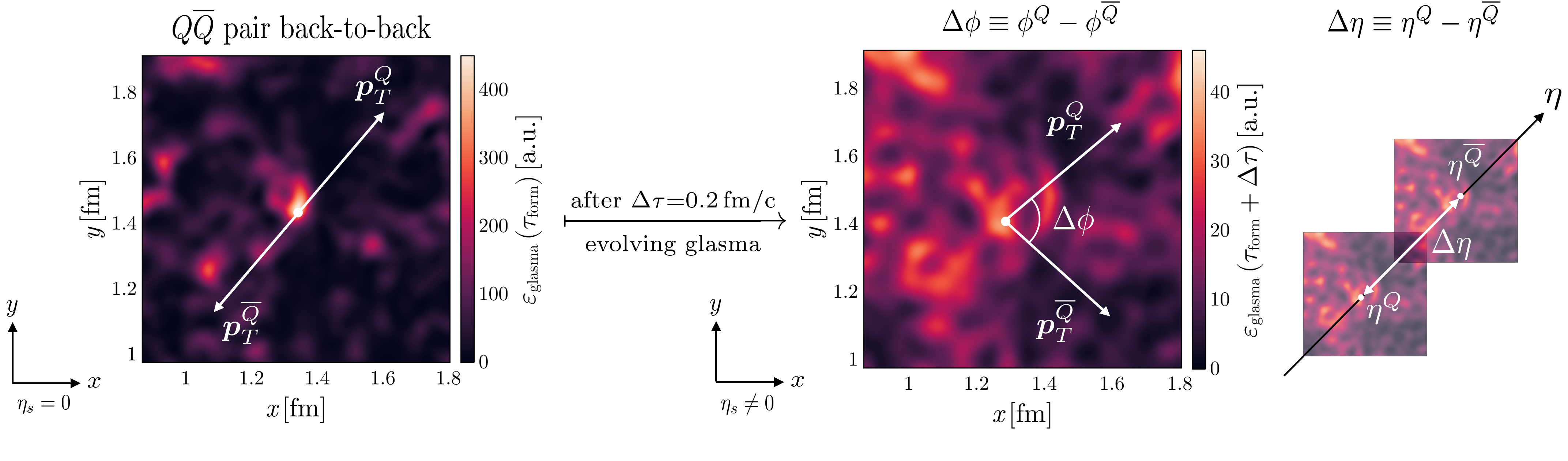}
\caption{Schematic representation of a $Q\overline{Q}$ pair produced with $\boldsymbol{p}_T^Q=-\boldsymbol{p}_T^{\overline{Q}}$ at $\eta=0$ (\textit{left panel}) and the evolved pair in the glasma fields, causing a change in $\Delta\phi$ (\textit{middle panel}) and $\Delta\eta$ (\textit{right panel}). The glasma energy density $\varepsilon_\mathrm{glasma}(\boldsymbol{x}_T)$ is shown in the background.}
\label{fig:qqbar}       
\end{figure*}

\begin{figure}
\centering
\sidecaption
\includegraphics[width=5cm,clip]{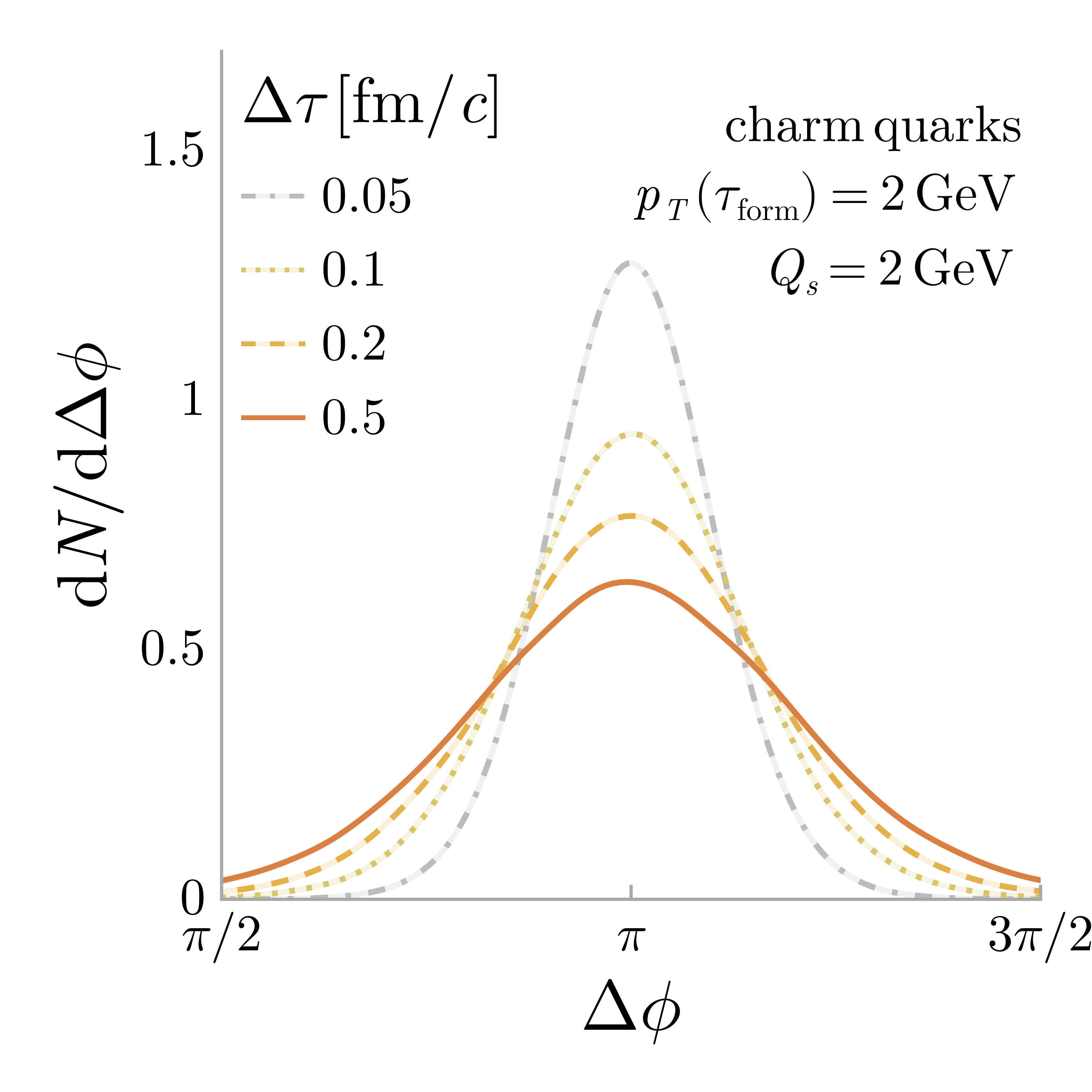}
\caption{Azimuthal correlation as a function of relative azimuthal angle $\Delta\phi$ for charm quarks with $p_T(\tau_\mathrm{form})=2\,\mathrm{GeV}$ in a glasma with $Q_s=2\,\mathrm{GeV}$ at different values of proper time $\tau$ (\textit{colored lines}). The correlation is extracted from large ensembles of test particles. The initial $\delta(\Delta\phi-\pi)$ peak quickly broadens due to the strong glasma fields. A similarly large decorrelation is obtained for beauty quarks in the glasma \cite{Avramescu:2024poa}.}
\label{fig:dndpt}   
\end{figure}

\noindent\textit{\textbf{Azimuthal correlation.}} An observable sensitive to the production of heavy quarks and thus the initial stage is the azimuthal correlation between $Q\overline{Q}$ pairs, shown in Fig.~\ref{fig:qqbar}. In vacuum, quark-antiquark pairs are produced back-to-back at leading-order (LO). As the quarks evolve through the glasma, their relative azimuthal angle $\Delta\phi$ and rapidity $\Delta\eta$ change. 

We calculate the azimuthal $\Delta\phi$ correlation of $c\overline{c}$ pairs in the glasma, represented in Fig.~\ref{fig:dndpt}. This relates to the experimentally measured $D\overline{D}$ angular correlations, planned for ALICE3~\cite{ALICE:2022wwr}. We observe that the initial back-to-back correlation is greatly reduced by the glasma fields. The decorrelation is driven by momentum kicks from the glasma. The glasma fields are strongest at early-times, causing the correlation width to rapidly increase.

\begin{figure*}
\centering
\vspace*{1cm}      
\includegraphics[width=0.95\textwidth,clip]{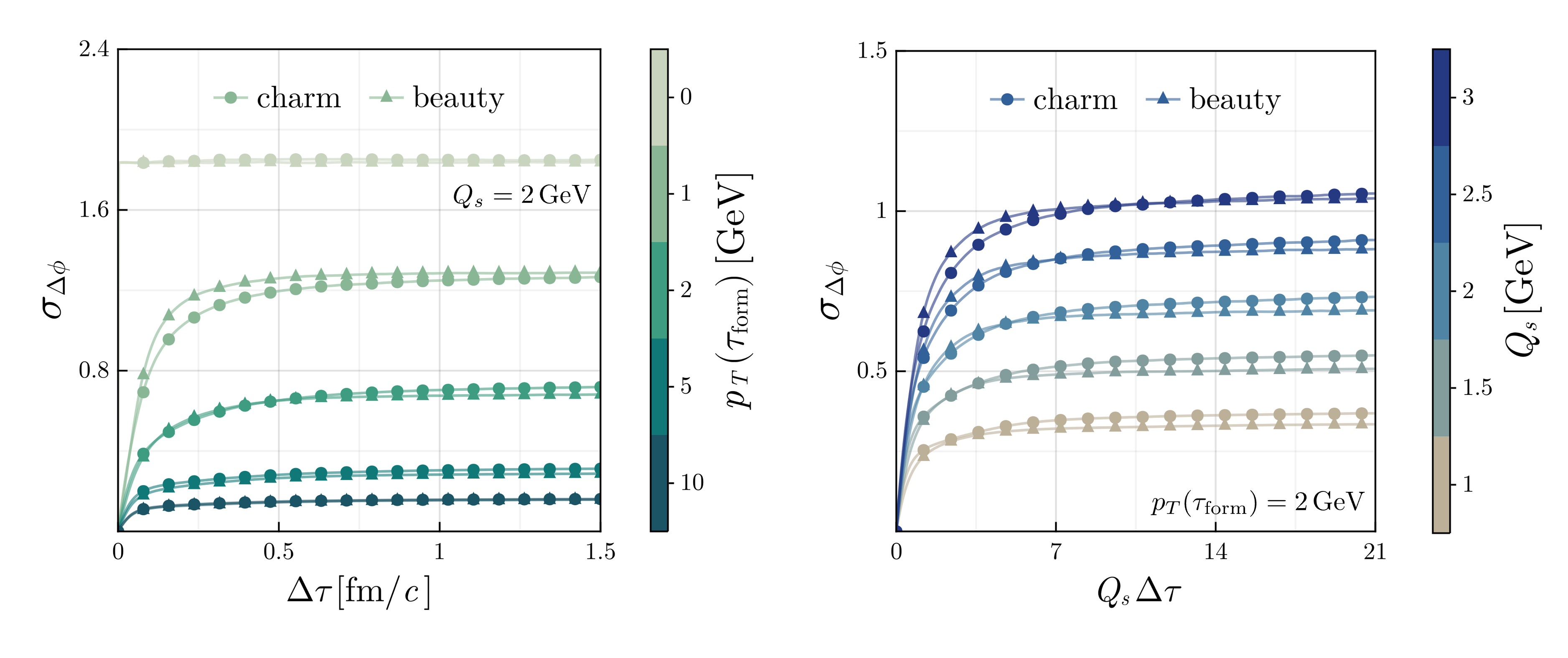}
\caption{Azimuthal decorrelation width $\sigma_{\Delta\phi}$ for charm (\textit{circle markers}) and beauty (\textit{triangle markers}) quarks as a function of $\Delta\tau$ for various $p_T$ (\textit{left panel}) and $Q_s$ (\textit{right panel}) values.}
\label{fig:sigma_dphi_pT_Qs_dep}       
\end{figure*}

To better quantify the decorrelation in the glasma, we extract the temporal evolution of the correlation width $\sigma_{\Delta\phi}$ as a function of quark $p_T$ and glasma $Q_s$, for both charm and beauty quarks, in Fig.~\ref{fig:sigma_dphi_pT_Qs_dep}. Quarks with lower $p_T$ experience the most decorrelation, while higher $p_T$ quarks better preserve the initial back-to-back correlation. Increasing the saturation momentum $Q_s$ and thus the strength of the glasma fields, enhances the decorrelation. We conclude that quark pairs with low initial $p_T$ in a glasma with high $Q_s$ (as at the LHC) experience significant azimuthal decorrelation. The value of the decorrelation in the glasma is of the same order of magnitude as that reported during the QGP phase \cite{Nahrgang:2013saa}.

\section{Summary} We developed a framework to simulate heavy quark transport in the glasma and obtained large transport coefficients. This framework enabled us to study the glasma's effects on observables such as the nuclear modification factor and heavy quark correlations. We found that the glasma alone moderately affects $R_{AA}$, while nPDF effects dominate. However, the glasma has a strong effect on $c\overline{c}$ azimuthal correlations, especially at low $p_T$, where a significant decorrelation is observed. In the near future, it is important to refine the current setup for these azimuthal correlations and compare with predictions during the QGP phase. As an outlook, we plan to explore energy loss in the glasma and potentially couple it to the kinetic theory phase. 

\noindent\textit{\textbf{Acknowledgments.}} This work was supported by the Research Council of Finland, the Centre of Excellence in Quark Matter (projects 346324 and 364191), and projects 338263, 346567, and 359902, and by the European Research Council (ERC, grant agreements ERC-2023-101123801 GlueSatLight and ERC-2018-ADG-835105 YoctoLHC). D.A. acknowledges the support of the Vilho, Yrj\"{o} and Kalle V\"{a}is\"{a}l\"{a} Foundation. D.M. acknowledges support from the Austrian Science Fund (FWF) projects P 34764 and P 34455.

\end{document}